\documentclass[aps,showkeys,twocolumn]{revtex4}
\usepackage{amsmath,amssymb}
\usepackage{graphics,graphicx}
\usepackage{dcolumn,bm}

\begin{document}
\title{Kinetic models for wealth exchange on directed networks}

\author{Arnab Chatterjee}%
\affiliation{Condensed Matter and
Statistical Physics Section,\\
The Abdus Salam International Centre for Theoretical Physics,
Strada Costiera 11, Trieste I-34014, Italy.}
\email[Email: ]{achatter@ictp.it}

\begin{abstract}
We propose some kinetic models of wealth exchange and investigate
their behavior on directed networks though numerical simulations.
We observe that network topology and directedness yields a variety
of interesting features in these models.
The nature of asset distribution
in such directed networks show varied results, the degree of asset
inequality increased with the degree of disorder in the graphs.
\end{abstract}
%\today


\maketitle
%%%%%%%%%%%%%%%%%%%%%%%%%%%%%%%%%%%%%%%%%%%%%%%%%%%%%%%%%%%%%%%%%%%
\section{Introduction}\label{sec:1}
%%%%%%%%%%%%%%%%%%%%%%%%%%%%%%%%%%%%%%%%%%%%%%%%%%%%%%%%%%%%%%%%%%%
The distribution of wealth among individuals in an economy has been an
important area of research in economics, for more than a hundred
years~\cite{Pareto:1897,Mandelbrot:1960,EWD05,ESTP}. 
The same is true for income distribution in any society.
Detailed analysis of the income distribution~\cite{EWD05,ESTP} so far
indicate
\begin{equation}
\label{par}
P(m) \sim 
\left\{ \begin{array}{lc}
m^\alpha \exp(-m/T) & \textrm{for} \ m < m_c,\\
m^{-(1+\nu)} \ \ \ \ & \textrm{for} \  m \ge m_c,
\end{array}\right.
\end{equation}
where $P$ denotes the number density of people with income 
or wealth $m$ and $\alpha$, $\nu$ denote exponents and $T$ denotes a scaling
factor. The power law in income and wealth distribution (for $m \ge m_c$) is
named after Pareto and the exponent $\nu$ is called the Pareto exponent.
A historical account of Pareto's data and that from recent sources can be
found in Ref.~\cite{Richmond:ESTP}.
The crossover point ($m_c$) is extracted from the crossover from a 
Gamma distribution form to the power law tail. One often fits the
region below $m_c$ to a log-normal form 
$\log P(m) \propto -(\log m)^2$.
Although this form is often preferred by economists, we think that the other
Gamma distribution form Eqn.~(\ref{par}) fits better with the data,
because of the remarkable fit with the Gibbs distribution in 
Ref.~\cite{Silva:2005,Willis:2004,DraguYakov01a}.

Considerable investigations revealed
that the tail of the income distribution indeed follows the above
mentioned behavior and the value of the Pareto exponent $\nu$ is generally
seen to vary between 1 and 
3~\cite{Dragulescu:2001,Levy:1997,Sinha:2006,Aoyama:2003,DiMatteo:2004,Clementi:2005,Ding:2007}. It is also
known that typically less than $10 \%$ of the population in any country
possesses about $40 \%$ of the total wealth of that country and they follow
the above law, while the rest of the low income population, 
follow a different distribution which is debated 
to be either 
Gibbs~\cite{Dragulescu:2001,Levy:1997,Aoyama:2003,marjit,Ispolatov:1998,Dragulescu:2000} 
or log-normal~\cite{DiMatteo:2004,Clementi:2005}.

The striking regularities %(see Fig.~\ref{fig:realdataset}) 
observed in the income distribution for different countries, have 
led to several new attempts at 
explaining them on theoretical grounds. Much of it is
from physicists' modeling of economic behavior in analogy with
large systems of interacting particles, as treated, e.g., in the kinetic
theory of gases. According to physicists, the
regular patterns observed in the income (and wealth) distribution may
be indicative of a natural law for the statistical properties of a
many-body dynamical system representing the entire set of economic 
interactions in a society, analogous to those previously derived for
gases and liquids. By viewing the economy as a thermodynamic system,
one can identify the income distribution with the distribution of
energy among the particles in a gas.
In particular, a class of kinetic exchange models have provided a simple
mechanism for understanding the unequal accumulation of assets.
Many of these models, while simple from the perspective of economics,
has the benefit of coming to grips with the key factor in socioeconomic
interactions that results in very different societies converging to
similar forms of unequal distribution of resources (see 
Refs.~\cite{EWD05,ESTP}, which consists of a collection of 
large number of technical papers in this field).

In this paper, we consider a new model of asset exchange on a directed
network and investigate the nature of wealth distribution, using
numerical simulations.
In Sec.~\ref{sec:idealgas} we review the salient features of the 
gas-like kinetic models. In Sec.~\ref{sec:dirnet}, we propose
the new model and present the results. We conclude with discussions
in Sec.~\ref{sec:disc}.

%%%%%%%%%%%%%%%%%%%%%%%%%%%%%%%%%%%%%%%%%%%%%%%%%%%%%%%%%%%%%%%%%%%%%%%%%%%
\section{Gas-like models}
\label{sec:idealgas}
%%%%%%%%%%%%%%%%%%%%%%%%%%%%%%%%%%%%%%%%%%%%%%%%%%%%%%%%%%%%%%%%%%%%%%%%%%%
In analogy to two-particle collision process which results in a change 
in their individual kinetic energy or momenta, income exchange 
models may be defined using two-agent interactions:
two randomly selected agents exchange 
money by some pre-defined mechanism. Assuming the exchange process 
does not depend on previous exchanges, the dynamics follows a 
Markovian process:
\begin{equation}
\left( \begin{array}{c}
m_i (t+1) \\
m_j (t+1)
\end{array}
\right) = {\mathcal M}
\left( \begin{array}{c}
m_i (t) \\
m_j (t)
\end{array}
\right)
\label{matrixM}
\end{equation}
where $m_i (t)$ is the income of (individual or corporate) 
agent $i$ at time $t$ and the 
collision matrix ${\mathcal M}$ defines the exchange mechanism.

In this class of models, one considers a closed economic system
where the total money $M$ and number of agents $N$ is fixed. 
This corresponds to no production or migration in the system 
where the only economic activity is confined to trading.
In any trading, a pair of traders $i$ and $j$ exchange their money
\cite{marjit,Ispolatov:1998,Dragulescu:2000,Chakraborti:2000}, 
such that their total money is locally conserved
and nobody ends up with negative money ($m_i(t) \ge 0$, i.e, debt not allowed):
\begin{equation}
\label{mdelm}
m_i(t+1) = m_i(t) + \Delta m; \  m_j(t+1) = m_j(t) - \Delta m.
\end{equation}
Time ($t$) changes by one unit after each trading.

The simplest model considers a random fraction of total money
to be shared~\cite{Dragulescu:2000}.
At steady-state ($t \rightarrow \infty$) money follows a Gibbs distribution:
$P(m)=(1/T)\exp(-m/T)$; $T=M/N$,
independent of the initial distribution.
This follows from the conservation of money and additivity of entropy:
\begin{equation}
\label{prob}
P(m_1)P(m_2)=P(m_1+m_2).
\end{equation}
This result is quite robust and 
is independent of the topology of the (undirected)
exchange space, be it regular lattice, fractal or 
small-world~\cite{Chakrabarti:2004}.

A saving propensity factor $\lambda$ was introduced in the random 
exchange model~\cite{Chakraborti:2000}, where each trader
at time $t$ saves a fraction $\lambda$ of its money $m_i(t)$ and trades
randomly with the rest:
\begin{equation}
\label{fmi}
m_i(t+1)=\lambda m_i(t) + \epsilon_{ij} \left[(1-\lambda)(m_i(t) + m_j(t))\right],
\end{equation}
\begin{equation}
\label{fmj}
m_j(t+1)=\lambda m_j(t) + (1-\epsilon_{ij}) \left[(1-\lambda)(m_i(t) + m_j(t))\right],
\end{equation}
$\epsilon_{ij}$ being a random fraction, coming from the stochastic nature
of the trading.

In this model (CC model hereafter), 
the steady state distribution $P(m)$ of money is 
decaying on both sides with the most-probable money per agent shifting away
from $m=0$ (for $\lambda =0$) to $M/N$ as 
$\lambda \to 1$ ~\cite{Chakraborti:2000}. 
This model has been understood to a certain 
extent~\cite{Patriarca:2004,Repetowicz:2005}
and argued to resemble a Gamma distribution~\cite{Patriarca:2004}.
But, the actual form of the distribution
for this model still remains to be found out.
It seems that a very similar model was proposed by 
Angle~\cite{Angle:1986,Angle:2006} several years back in sociology journals.
The numerical simulation results of Angle's model 
fit well to Gamma distributions.

Empirical
observations in homogeneous groups of individuals as in waged income
of factory labourers in UK and USA~\cite{Willis:2004}
and data from population survey in USA among students of different school
and colleges produce similar distributions~\cite{Angle:2006}. 
This is a simple case where a homogeneous population 
(say, characterized by a unique value of $\lambda$) has been identified.

In a real society or economy, 
saving $\lambda$ is a very inhomogeneous parameter.
The evolution of money in a corresponding trading model (CCM model hereafter) 
can be written as~\cite{Chatterjee:2004}:
\begin{equation}
\label{mi}
m_i(t+1)=\lambda_i m_i(t) + \epsilon_{ij} \left[(1-\lambda_i)m_i(t) + (1-\lambda_j)m_j(t)\right],
\end{equation}
\begin{equation}
\label{mj}
m_j(t+1)=\lambda_j m_j(t) + (1-\epsilon_{ij}) \left[(1-\lambda_i)m_i(t) + (1-\lambda_j)m_j(t)\right]
\end{equation}
The trading rules are same as CC model, except that
$\lambda_{i}$ and $\lambda_{j}$, the saving propensities of agents 
$i$ and $j$, are different. The agents have fixed (over time) saving
propensities, distributed independently, randomly as $\Lambda(\lambda)$,
such that $\Lambda(\lambda)$ is non-vanishing as $\lambda \to 1$,
$\lambda_i$ value is quenched
for each agent ($\lambda_i$ are independent of trading or $t$).
The actual asset distribution $P(m)$ in such a model will depend on the 
form of $\Lambda(\lambda)$, but for all of them the asymptotic form of 
the distribution will become Pareto-like: $P(m) \sim m^{-(1+\nu)}$;
$\nu = 1$ for $m \to \infty$. This is valid for all such distributions,
unless $\Lambda(\lambda) \propto (1-\lambda)^{\delta}$, when
$P(m) \sim m^{-(2+\delta)}$~\cite{Chatterjee:2004,Mohanty:2006,Chatterjee:rev}.
In the CCM model, agents with higher saving propensity tend to hold
higher average wealth, which is justified by the fact that the saving
propensity in the rich population is always high~\cite{Dynan:2004}.
Analytical understanding of CCM model has been possible until now
under certain approximations~\cite{Chatterjee:2005}, and mean-field 
theory~\cite{Mohanty:2006}.
Unlike the above models with savings as a quenched disorder, one can
also consider savings as an annealed variable and still derive
a power law distribution in wealth~\cite{ecoanneal}.

%%%%%%%%%%%%%%%%%%%%%%%%%%%%%%%%%%%%%%%%%%%%%%%%%%%%%%%%%%%%%%%%%%%%%%%%%%%%
\section{Models on directed networks}
\label{sec:dirnet}
%%%%%%%%%%%%%%%%%%%%%%%%%%%%%%%%%%%%%%%%%%%%%%%%%%%%%%%%%%%%%%%%%%%%%%%%%%%
%%%%%%%%%%%%%%%%%%%%%%%%%%%%%%%%%%%%%%%%%%%%%%%%%%%%%%%%%%%%%%%%5
\begin{figure}[b]
\centering \includegraphics[width=9.0cm]{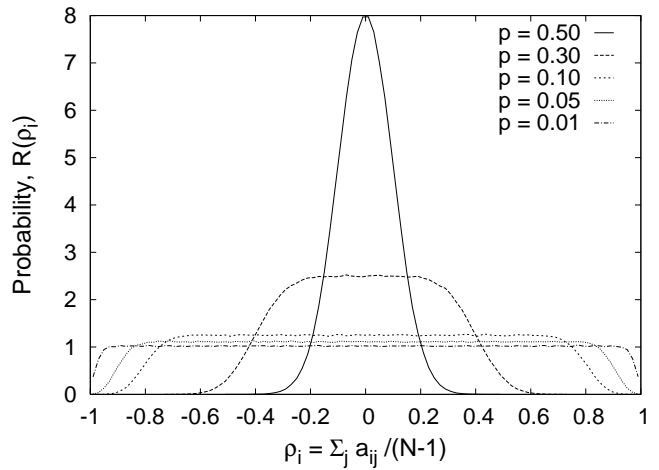}
\caption{
The distribution $R(\rho_i)$ of the link disorder $\rho_i$ 
for a directed network with different values of 
$p = 0.50, 0.30, 0.10, 0.05, 0.01$ for a system of $N=100$ nodes, 
obtained by numerical simulation, averaged over $10^4$ realizations.
}
\label{fig:dirnet:io}
\end{figure}
%%%%%%%%%%%%%%%%%%%%%%%%%%%%%%%%%%%%%%%%%%%%%%%%%%%%%%%%%%%%%%%%5
The topology of exchange space in a real society is quite complicated. 
A mean field scenario does not take into account the constraints on the 
flow of money or wealth. In other words, there are strong notions of 
directionality and sometimes, hierarchy in the underlying network,
where money is preferentially transeferred in certain direction
that others, contributing in irreversible flow of money.
A way to imitate this is to consider wealth exchange models
on a directed network~\cite{Wasserman:1994,Albert:2002,Dorogovtsev:2003}.
There have been previous attempts to obtain the same using the physics
of networks~\cite{Hu:2006,Hu:2007,Garlaschelli:2008}.
In this section, we would propose suitable toy models and present some 
interesting observations.

%%%%%%%%%%%%%%%%%%%%%%%%%%%%%%%%%%%%%%%%%%%%%%%%%%%%%%%%%%%%%%%%%%%%%%%%%%%%
\subsection{The Model}
\label{subsec:model}
%%%%%%%%%%%%%%%%%%%%%%%%%%%%%%%%%%%%%%%%%%%%%%%%%%%%%%%%%%%%%%%%%%%%%%%%%%%
We consider $N$ agents, each at a separate node, 
each of which are connected to the rest $N-1$ by
directed links. The directionality of the links denote 
the direction of flow of wealth in this fully connected network.
The directed network parametrized by $p$ is constructed in the following way:

(a) there are no self links, so that the adjacency matrix 
$\cal{A}$~\cite{Albert:2002,Dorogovtsev:2003} has
diagonal elements $a_{ii} = 0$ for all sites $i$.

(b) for each matrix element $a_{ij}$, $i \ne j$, we call a random number $r$.
$a_{ij} = +1$ if $r < p$ and $a_{ij} = -1$ otherwise. Also $a_{ji} = -a_{ij}$.
Thus we have $N(N-1)/2$ such calls of $r$.
$a_{ij} = +1$ denotes a link directed from $i$ to $j$ and 
$a_{ij} = -1$ denotes a link directed from $j$ to $i$.

We define $\rho_i = \frac{1}{N-1}\sum_j a_{ij}$ 
as a measure of link disorder at 
the site $i$. $\sum_j$ denotes the sum over all $N-1$ sites $j$ linked to $i$. 
Thus, $\rho_i = 1$ is a node for which all links are outgoing and
$\rho_i = -1$ is a node for which all links are incoming.
The parameter $p$ has a symmetry about $0.5$ and 
the distribution $R(\rho_i)$ is also symmetric about $0$,
which is, in fact, a consequence of the conservation of the number of incoming
and outgoing links. A network with $p=0.5$ has the lowest degree of
disorder, given by a narrow distribution $R(\rho)$ of $\rho$, 
around $\rho = 0$ (see Fig.~\ref{fig:dirnet:io}).
This means that almost all nodes have more or less equal number of incoming and 
outgoing links. On the other extreme, $p=0.01$ is a network which has a small
but finite number of nodes where most links are incoming/outgoing, thus giving
rise to a very wide distribution of link disorder $R(\rho)$.
%%%%%%%%%%%%%%%%%%%%%%%%%%%%%%%%%%%%%%%%%%%%%%%%%%%%%%%%%%%%%%%%5
\begin{figure*}%[t]
\centering \includegraphics[width=8.8cm]{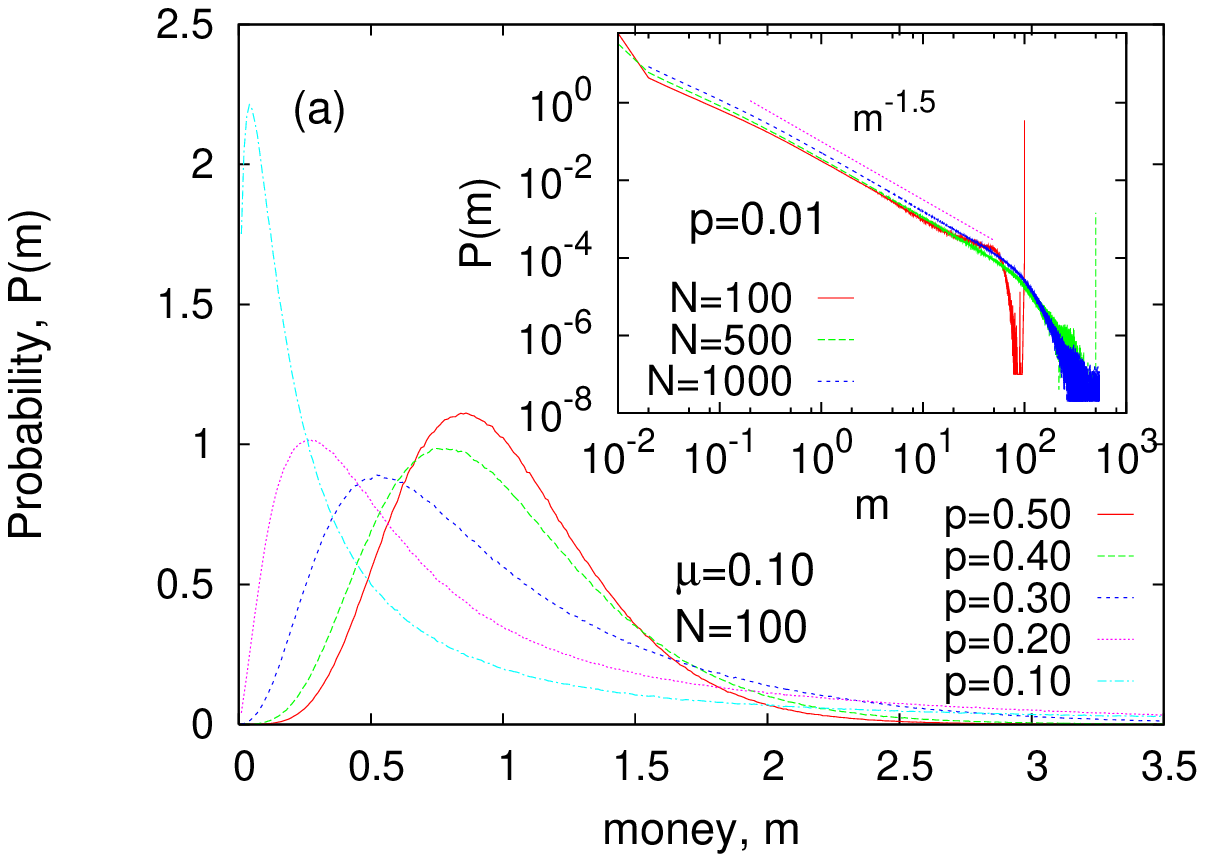}
\centering \includegraphics[width=8.8cm]{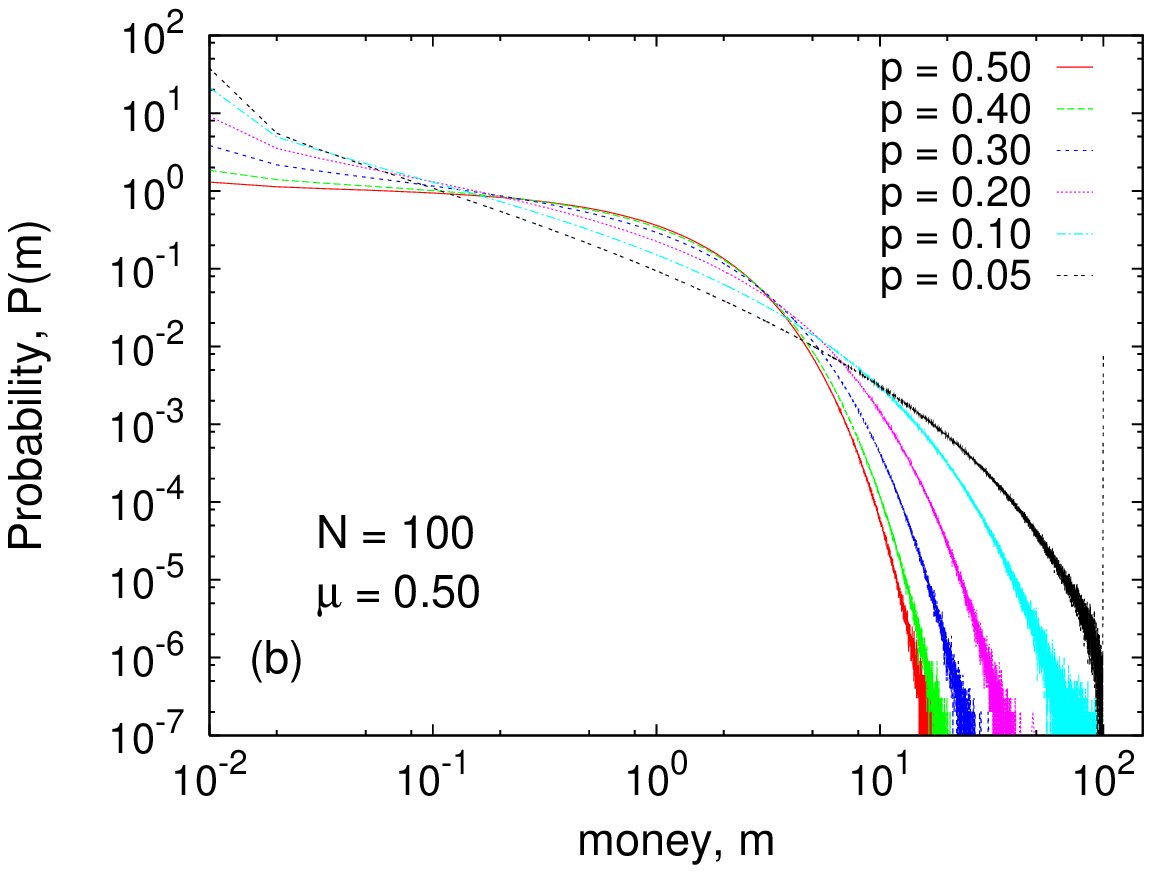}
\centering \includegraphics[width=8.8cm]{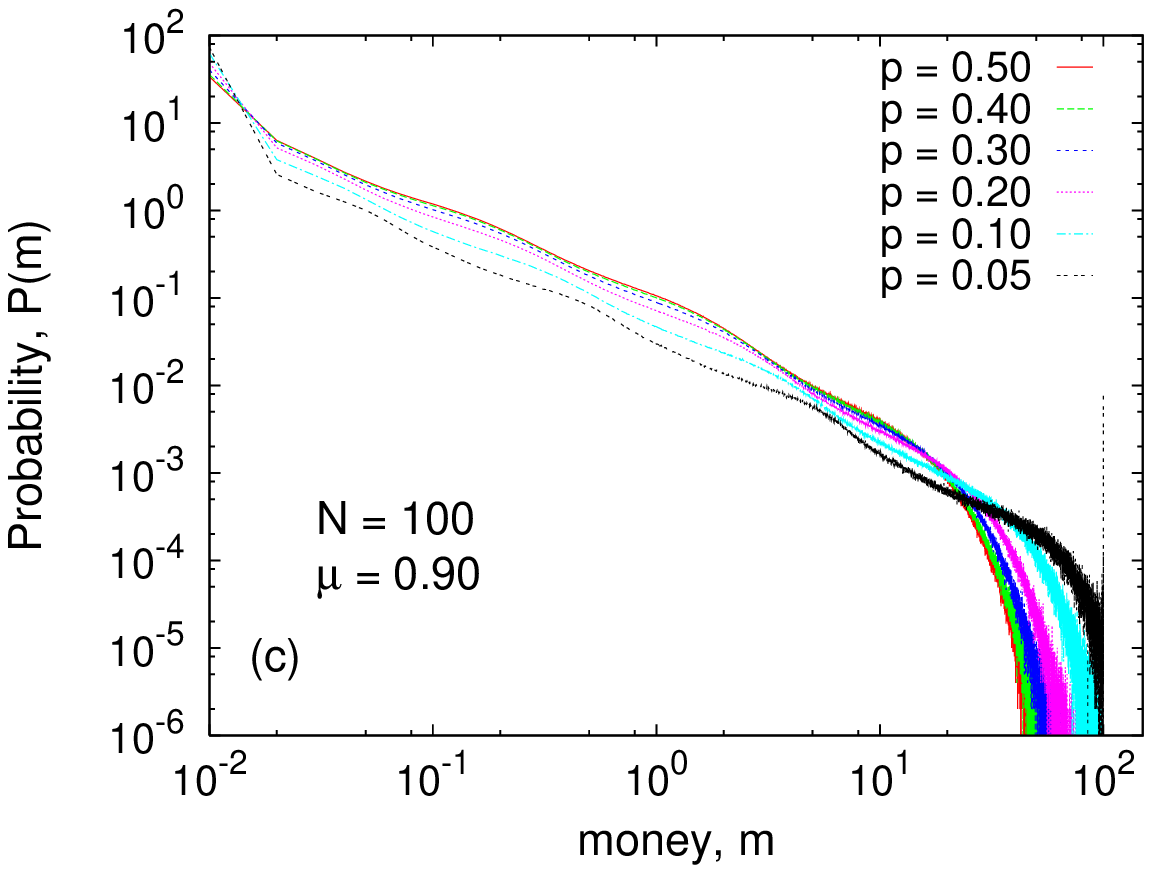}
\centering \includegraphics[width=8.8cm]{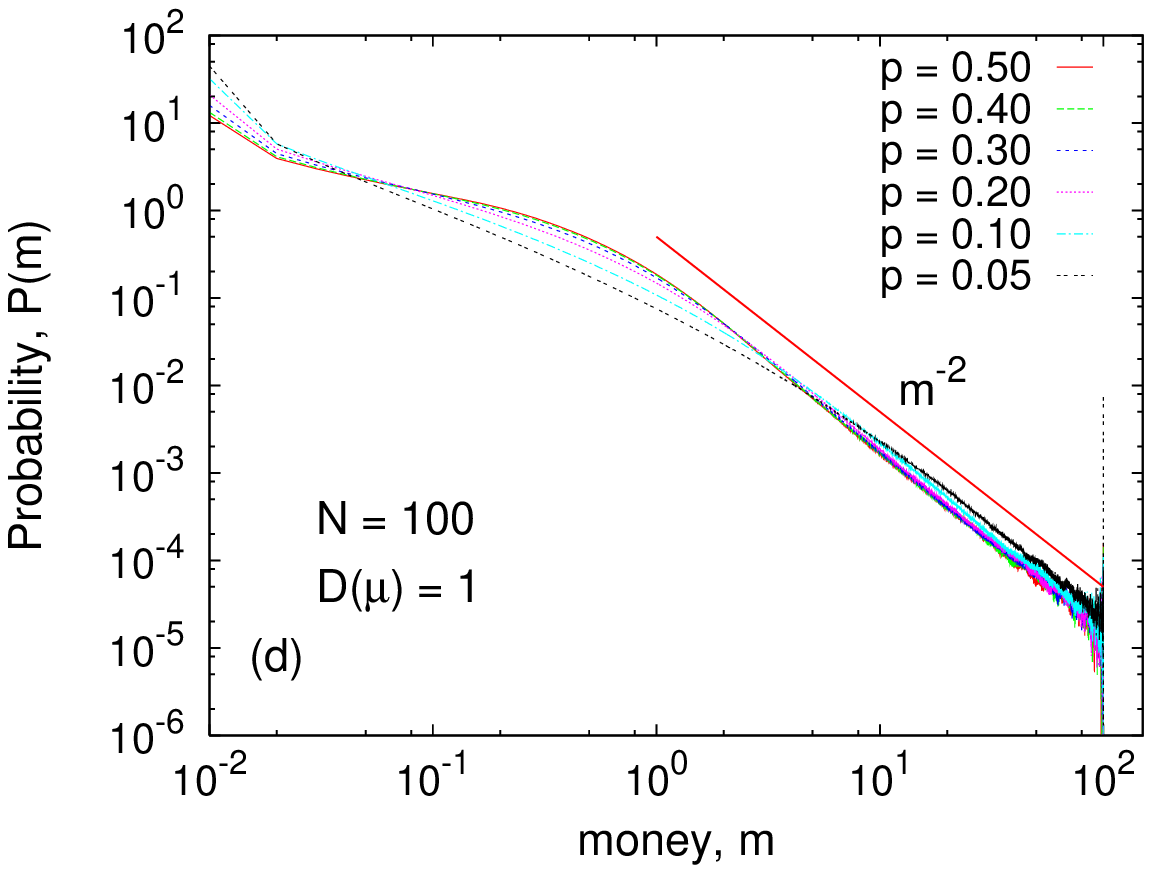}
\caption{
The steady state distribution $P(m)$ of money $m$ 
for directed networks characterized by different values of 
$p$. 
(a) For $\mu = 0.1$, the inset shows $P(m)$ for $p = 0.01$ for
$N=100, 500, 1000$, and the power law $m^{-1.5}$ is also indicated. 
(b) For $\mu = 0.5$ and (c) $\mu = 0.9$.
(d) shows the plots for uniform, random distributed $\mu$,
$D(\mu) = 1$, and also a guide to the power law $m^{-2}$.
The data are obtained by numerical simulation, for a system of $N=100$ nodes, 
averaged over $10^3$ realizations in the steady state and over $10^4$ initial
configurations. The average money $M/N$ is $1$.
}
\label{fig:dn:pm}
\end{figure*}
%%%%%%%%%%%%%%%%%%%%%%%%%%%%%%%%%%%%%%%%%%%%%%%%%%%%%%%%%%%%%%%%5
The rules of exchange are as follows:
If $a_{ij} = -1$,
\begin{eqnarray}
\label{eqn:dir-ex1}
m_i(t+1) &=& m_i(t) + \mu_j m_j(t) \nonumber\\
m_j(t+1) &=& m_j(t) - \mu_j m_j(t) \nonumber\\
\end{eqnarray}
else, $a_{ij} = +1$,
\begin{eqnarray}
\label{eqn:dir-ex2}
m_i(t+1) &=& m_i(t) - \mu_i m_i(t) \nonumber\\
m_j(t+1) &=& m_j(t) + \mu_i m_i(t). \nonumber\\
\end{eqnarray}
where $0 < \mu_i < 1$ is a `transfer fraction' associated with the $i$th agent,
and $m_i(t)$ is the money of agent $i$ (or, money at node $i$) at time $t$.
The total money in the system is conserved, no money is created or destroyed, 
as is evident from eqn.~(\ref{eqn:dir-ex1}) and eqn.~(\ref{eqn:dir-ex2}). 
If there is a link from $j$ to $i$, the node $i$ gains $\mu_j$ fraction
of $j$th agent's money, which, of course agent $j$ loses. Otherwise,
if there is a link from $i$ to $j$, the node $j$ gains $\mu_i$ fraction
of $i$th agent's money, which, of course agent $i$ loses.
In the Monte Carlo simulations, one assigns random amount of money to agents
to start with, such that the average money $M/N=1$. 
A pair of agents (nodes) are chosen at random,
and depending on the directionality of the link between them (the sign of
$a_{ij}$), the relevant rule, eqn.~(\ref{eqn:dir-ex1})
or eqn.~(\ref{eqn:dir-ex2}) is chosen. 
This is repeated until a steady state is
reached and the distribution of money does not change in time. 
The distribution of money $P(m)$ is obtained by averaging over 
several ensembles (different random initial distribution of money).

This model is different from the CC and CCM models, but one can relate the
transfer fraction $\mu$ analogous to $\lambda$ in the CC and CCM models.

%%%%%%%%%%%%%%%%%%%%%%%%%%%%%%%%%%%%%%%%%%%%%%%%%%%%%%%%%%%%%%%%%%%%%%%%%%%%
\subsection{Model with uniform $\mu$}
\label{subsec:fix-mu}
%%%%%%%%%%%%%%%%%%%%%%%%%%%%%%%%%%%%%%%%%%%%%%%%%%%%%%%%%%%%%%%%%%%%%%%%%%%
First we discuss the case of homogeneous agents, i.e, when 
all agents $i$ have $\mu_i = \mu$.
The $\mu=0$ limit is trivial, as the system does not perform any dynamics.
Fig.~\ref{fig:dn:pm}(a) shows the steady state distribution $P(m)$ of
money $m$ for $\mu=0.1$ for different values of network disorder $p$.
In general, the distribution of money has a most probable value,
which shifts monotonically from about $0.85$ for $p=0.5$ to $0$ as $p \to 0$.
$P(m)$ has an exponential tail, but a power law region develops as $p \to 0$
in between the most probable peak and the exponential cut-off, which fits 
approximately to $m^{-1.5}$.
For the limit $p \to 0$, condensation of wealth at the node(s) with strong
disorder ($\rho \to 1$) is apparent from the single, 
isolated data point at the $m_{max}=M$
end (see inset of Fig.~\ref{fig:dn:pm}(a), for $p=0.01$).
There is a strong finite size effect involved in this behavior.
To emphasize this, we plot $P(m)$ for $p=0.01$ for $N=100,500$ and 
$1000$ (inset of Fig.~\ref{fig:dn:pm}(a)).
While $N=100$ and $N=500$ does show the isolated data point, it is
absent for $N=1000$. This indicates that this behavior is absent 
for infinite systems for $p \to 0$.
For larger values of $p$, the distribution resembles
Gamma distributions, as in the CC model.
For $\mu=0.5$, the most-probable value of $P(m)$ is always at $0$ (see
Fig.~\ref{fig:dn:pm}(b)). For weak disorder ($p=0.5$), $P(m)$ is exponential,
but it goes to a wider distribution as one goes to higher disorder ($p \to 0$).
The condensation of wealth at node(s) with high value of $\rho$ ($\rho \to 1)$ 
is again apparent from the single, isolated data point at the $m_{max}=M$
end (see Fig.~\ref{fig:dn:pm}(b), for $p=0.05$).
For $\mu=0.9$, $P(m)$ is always decaying, with a wide distribution upto
$m_{max}=M$ (see Fig.~\ref{fig:dn:pm}(c)). As like previous plots, the
condensation of wealth at node(s) with high value of $\rho$ is visible:
see plot for $p=0.05$ in Fig.~\ref{fig:dn:pm}(c).
A common feature for the curves for all values of $p$ is that, $P(m)$
exhibits log-periodic oscillations, while resembling roughly a power-law
decay. Another important feature is that $P(m) \to N$ for $m \to 0$,
which indicates that money is distributed in a very small fraction of 
nodes, while most nodes have almost no money at a given instance.

For a particular value of the network disorder $p$, the wealth distribution
$P(m)$ becomes more and more `fat tailed' as $\mu$ is increased.
This is in contrast to what is observed in CC model~\cite{Chakraborti:2000}
where $P(m)$ organizes to a narrower distribution as $\lambda$ increases.
%%%%%%%%%%%%%%%%%%%%%%%%%%%%%%%%%%%%%%%%%%%%%%%%%%%%%%%%%%%%%%%%%%%%%%%%%%%%
\subsection{Model with distributed $\mu$}
\label{subsec:dis-mu}
%%%%%%%%%%%%%%%%%%%%%%%%%%%%%%%%%%%%%%%%%%%%%%%%%%%%%%%%%%%%%%%%%%%%%%%%%%%
We now consider the case when agents $i$ have have different values of $\mu_i$,
which do not change in time. This is a case of heterogeneous agents where
the heterogeneity can be viewed as a `quenched disorder'.
We consider a random uniform distribution of $\mu$, i.e,
$D(\mu) = 1$ in $0< \mu < 1$. 
This is the case analogous to the CCM model~\cite{Chatterjee:2004}.
Fig.~\ref{fig:dn:pm}(d) shows the plots
of the money distribution $P(m)$ for different values of network disorder $p$. 
All curves have a power law tail, resembling a $m^{-2}$ variation. However,
the effect of the topology of the underlying network are visible:
For strong disorder in topology $p=0.05$, condensation of wealth
at node(s) with high value of $\rho$ ($\rho \to 1)$ is also apparent 
from the single, isolated data point at the $m_{max}=M$ end.
Further investigations also indicate that the power law exponent
is similarly related to the distribution of `transfer fraction' $\mu$,
as one observes in the CCM model~\cite{Chatterjee:2004,Mohanty:2006}, i.e, 
$P(m) \sim m^{-2}$ for most distributions, while one can obtain 
$P(m) \sim m^{-(2+\delta)}$ if $D(\mu) \propto (1-\mu)^{\delta}$.

%%%%%%%%%%%%%%%%%%%%%%%%%%%%%%%%%%%%%%%%%%%%%%%%%%%%%%%%%%%%%%%%%%%%%%%%%%%%
\section{Discussions}
\label{sec:disc}
%%%%%%%%%%%%%%%%%%%%%%%%%%%%%%%%%%%%%%%%%%%%%%%%%%%%%%%%%%%%%%%%%%%%%%%%%%%
The empirical data for income and wealth distribution in many countries
are now available, and they reflect a particular robust 
pattern: The bulk (about 90\%) of the distribution resemble the century-old 
Gibbs distribution of energy for an ideal gas, while there are evidences 
of considerable deviation in the low income as well as high income ranges.
The high income range data (for 5-10\% of the population in any country) 
fits to a power law tail, known after Pareto, and the value of the 
(power law) exponent ranges between 1-3 and depends on the individual
make-up of the economy of the society or country.

The analogy with a gas like many-body system has led
to the formulation of the models of markets.
The random scattering-like dynamics of money (and wealth) in a closed 
trading market, in analogy with energy conserved exchange models,
reveals interesting features.
Self-organization is a key emerging feature of these kinetic exchange
models when saving factors are introduced.
These models~\cite{Chakraborti:2000,Chatterjee:2004}
produce asset distributions resembling that observed in reality,
and are quite well studied now~\cite{Chatterjee:rev}. Empirical
observations in homogeneous groups of individuals as in waged income
of factory labourers in UK and USA~\cite{Willis:2004}
and data from population survey in USA among students of different school
and colleges produce similar distributions~\cite{Angle:2006}.
There have also been other models~\cite{Sinha:2003} which
uses the kinetic exchange approach.

Heterogeneity is an inherent feature of real networked economy,
mostly observed with with behavior of agents contributing to
irreversible flow of money. A simple way of looking into it
is to consider heterogeneity in the interaction topology for
the agents.
In all earlier studies, the effect of the topology of the space
on which the kinetic exchanges took place, was overlooked. Most of
the models were either on regular, undirected lattices, or on undirected 
graphs. Our study concentrates on the effect of a directed network
as the exchange space for such kinetic exchanges.
We consider a simple, fully connected, directed network.
A notion of disorder $\rho$ in the directedness of the graph was introduced,
in analogy to real world networks, which depend on the number
of incoming and outgoing links at a particular node. 
The distribution $R(\rho)$ characterizes a particular network.
In analogy to saving propensity $\lambda$ in CC and CCM 
models~\cite{Chakraborti:2000,Chatterjee:2004}, we define a `transfer fraction'
$\mu$. We consider two distinct models: (i) with uniform 
$\mu$ (Sec.~\ref{subsec:fix-mu}), and 
(ii) with distributed $\mu$ (Sec.~\ref{subsec:dis-mu}). 
We employ numerical simulations to study the above models.
For uniform $\mu$, when the transfer fraction $\mu$ is small, 
low degree of disorder in the network
produced money distribution $P(m)$ with most-probable money at 
finite values of $m$ similar to Gamma distribution observed
in CC model~\cite{Chakraborti:2000}, 
while high disorder produced a wider distribution 
of $P(m)$ (Fig.~\ref{fig:dn:pm}(a)). At intermediate ranges of  $\mu$,
$P(m)$ has its most-probable value always at $m \to 0$ 
(Fig.~\ref{fig:dn:pm}(b)), while for high values of $\mu$,
$P(m)$ has a wide distribution with log-periodic oscillations
(Fig.~\ref{fig:dn:pm}(c)).
For the model with $\mu$ distributed randomly and uniformly ($D(\mu)=1$), 
money distribution exhibits power law
$P(m) \sim m^{-2}$ for large money $m$ (Fig.~\ref{fig:dn:pm}(d)).
The common feature of all these cases was a very wide distribution $P(m)$
of money, when the network disorder is very strong (see plots for $p=0.05$
in Fig.~\ref{fig:dn:pm}), accompanied by a signature of accumulation of money
at the nodes/with the agents having strongest disorder, which is observed
in the data point at $m_{max}=M$.
However, this effect is more pronounced in smaller system sizes,
while it may disappear in the thermodynamic limit 
(see inset of Fig.~\ref{fig:dn:pm}(a)).

A simple model of similar exchanges have been studied 
before~\cite{Dragulescu:2000}. The case corresponds to the choice $p=0.5$
for our model. Moreover, the transferred money was independent of $m$ and 
not proportional to $\mu$. Money distribution was found to be exponential
despite explicit violation of time-reversal symmetry.
In an earlier paper~\cite{Ispolatov:1998}, results for multiplicative
random exchanges correspond to certain cases of the present paper.
It was found that $P(m)$ vanishes at $m=0$ for $\mu < 0.5$
and diverges for $\mu > 0.5$, and is exponential for $\mu=0.5$.
These results correspond to that of our model for $p=0.5$ 
(Sec.~\ref{subsec:fix-mu}).

We have mostly used the terms `money' and `wealth' interchangeably,
treating the models in terms of only one quantity, namely `money' that
is exchanged. Of course, wealth does not comprise of (paper)
money only, and there have been studies distinguishing these 
two~\cite{Chatterjee:rev,Silver:2002,Ausloos:2007}.

Study of such simple models here give some insight into the possible
emergence of self organizations in such markets, evolution of the steady 
state distribution, emergence of Gamma-like distribution for the bulk
and of the power law tail, as in the empirically observed distributions. 
The study of models on directed graphs gives us new insights: (i) one can
obtain wide asset distribution even in homogeneous populations (as in
uniform $\mu$), (ii) the power law exponent for the distributed $\mu$ case
behaves in the same manner as in CCM model, and
(iii) strong disorder in topology of the underlying directed network
produces accumulation (at $\rho \to 1$) or the opposite effect
(at $\rho \to -1$) of wealth, thus giving rise to a higher degree
of wealth inequality in the system.

These model studies also indicate
the appearance of self-organization, and the self-organized 
criticality~\cite{Bak:1997} in particular, in the simple models.
These have prospective applications in other spheres of social science, as 
in application in policy making and taxation,
and also physical sciences, in designing desired energy 
spectrum for different types of chemical reactions~\cite{Scafetta:2007}.

%%%%%%%%%%%%%%%%%%%%%%%%%%%%%%%%%%%%%%%%%%%%%%%%%%%%%%%%%%%%%%%%%%%%
\begin{acknowledgments}
The author thanks 
B.~K.~Chakrabarti, M.~Marsili and P. Sen
for some useful comments and discussions.
The author is also thankful to the anonymous referee for
pointing out correspondences with previous works.
The author was supported by the ComplexMarkets E.U. STREP project  
516446 under FP6-2003-NEST-PATH-1.
\end{acknowledgments}
%%%%%%%%%%%%%%%%%%%%%%%%%%%%%%%%%%%%%%%%%%%%%%%%%%%%%%%%%%%%%%%%%%%%

%%%%%%%%%%%%%%%%%%%%%%%%%%%%%%%%%%%%%%%%%%%%%%%%%%%%%%%%%%%%%%%%%%%%

%%%%%%%%%%%%%%%%%%%%%%%%%%%%%%%%%%%%%%%%%%%%%%%%%%%%%%%%%%%%%%%%%%%
\end{document}